\documentclass[10pt,conference,fleqn,]{IEEEtran}
\IEEEoverridecommandlockouts
\usepackage[utf8]{inputenc}
\usepackage{listings}
\usepackage{color}
\usepackage{xspace}
\usepackage[usenames, dvipsnames]{xcolor}
\usepackage{graphicx}
\usepackage[normalem]{ulem}
\usepackage[many]{tcolorbox}
\usepackage{enumitem}
\usepackage{amsmath}
\usepackage{bm}
\usepackage{setspace}
\usepackage[ruled,lined,linesnumbered,vlined,algo2e]{algorithm2e}
\usepackage{cite}
\usepackage{soul}
\usepackage{url}
\usepackage{caption}
\usepackage{ulem}
\usepackage{booktabs}
\usepackage{hyperref} 
\usepackage{cleveref}
\usepackage{balance}

\makeatletter
\newcommand{\linebreakand}{%
  \end{@IEEEauthorhalign}
  \hfill\mbox{}\par
  \mbox{}\hfill\begin{@IEEEauthorhalign}
}
\makeatother

\def\BibTeX{{\rm B\kern-.05em{\sc i\kern-.025em b}\kern-.08em
    T\kern-.1667em\lower.7ex\hbox{E}\kern-.125emX}}
\makeatletter

\usepackage{makecell}
\usepackage{threeparttable}
\makeatletter  
\newif\if@restonecol  
\renewcommand\footnoterule{%
	\kern-3\p@
	\hrule\@width\columnwidth
	\kern2.6\p@}

\author{
\IEEEauthorblockN{Lyuye Zhang\IEEEauthorrefmark{1},
Chengwei Liu\IEEEauthorrefmark{1},
Zhengzi Xu\IEEEauthorrefmark{1},
Sen Chen\IEEEauthorrefmark{2},
Lingling Fan\IEEEauthorrefmark{3},
Lida Zhao\IEEEauthorrefmark{1},
Jiahui Wu\IEEEauthorrefmark{1},
Yang Liu\IEEEauthorrefmark{1},
}
\thanks{Chengwei Liu and Sen Chen are the corresponding authors.}
\IEEEauthorblockA{\IEEEauthorrefmark{1}School of Computer Science and Engineering\\
Nanyang Technological University, Singapore\\
zh0004ye@e.ntu.edu.sg, chengwei001@e.ntu.edu.sg, zhengzi.xu@ntu.edu.sg, \\lida001@e.ntu.edu.sg, jiahui004@e.ntu.edu.sg, yangliu@ntu.edu.sg}
\IEEEauthorblockA{\IEEEauthorrefmark{2}College of Intelligence and Computing, Tianjin University, Tianjin, China\\
senchen@tju.edu.cn}
\IEEEauthorblockA{\IEEEauthorrefmark{3}College of Cyber Science, Nankai University, Tianjin, China\\
linglingfan@nankai.edu.cn}
}

\newcommand\xleftrightarrow[2][]{%
  \ext@arrow 9999{\longleftrightarrowfill@}{#1}{#2}}
\newcommand\longleftrightarrowfill@{%
  \arrowfill@\leftarrow\relbar\rightarrow}
\makeatother
\usepackage{tikz}
\newcommand*\emptycirc[1][1ex]{\tikz\draw (0,0) circle (#1);} 
\newcommand*\halfcirc[1][1ex]{%
  \begin{tikzpicture}
  \draw[fill] (0,0)-- (90:#1) arc (90:270:#1) -- cycle ;
  \draw (0,0) circle (#1);
  \end{tikzpicture}}
\newcommand*\fullcirc[1][1ex]{\tikz\fill (0,0) circle (#1);} 

\setlist[itemize]{noitemsep, topsep=0pt}

\AtBeginDocument{%
  \providecommand\BibTeX{{%
    \normalfont B\kern-0.5em{\scshape i\kern-0.25em b}\kern-0.8em\TeX}}}

\newcommand{\tool}{\textsc{\textsc{Coral}}\xspace}
\newcommand{\dpb}{Dependabot\xspace}
\newcommand{\scantist}{Com B\xspace}
\newcommand{\depg}{DG\xspace}
\newcommand{\depgs}{DGs\xspace}
\newcommand{\snyk}{Com A\xspace}
\newcommand{\steady}{Steady\xspace}

\newcommand{\rip}{\textit{ripple effects}\xspace}
\newcommand{\ma}{\textit{Major}\xspace}
\newcommand{\mi}{\textit{Minor}\xspace}
\newcommand{\pa}{\textit{Patch}\xspace}
\newcommand{\bla}{\textit{Baseline A}\xspace}
\newcommand{\blb}{\textit{Baseline B}\xspace}
\newcommand{\blc}{\textit{Baseline C}\xspace}

\begin{document}
\title{Compatible Remediation on Vulnerabilities from Third-Party Libraries for Java Projects }

\maketitle

\begin{abstract}
With the increasing disclosure of vulnerabilities in open-source software, software composition analysis (SCA) has been widely applied to reveal third-party libraries and the associated vulnerabilities in 
software projects. Beyond the revelation, SCA tools adopt various remediation strategies to fix vulnerabilities, the quality of which varies
substantially. However, ineffective remediation could induce side effects, such as compilation failures, which impede acceptance by users. 
According to our studies, existing SCA tools could not correctly handle the concerns of users regarding the compatibility of remediated projects. To this end, we propose \underline{Co}mpatible \underline{R}emediation of Third-p\underline{a}rty \underline{l}ibraries (\tool) for Maven projects to fix vulnerabilities without breaking the projects. The evaluation proved that \tool not only fixed $87.56\%$ of vulnerabilities which outperformed other tools (best $75.32\%$) and achieved a $98.67\%$ successful compilation rate and a $92.96\%$ successful unit test rate. Furthermore, we found that $78.45\%$ of vulnerabilities in popular Maven projects could be fixed without breaking the compilation, and the rest of the vulnerabilities ($21.55\%$) could either be fixed by upgrades that break the compilations or even be impossible to fix by upgrading.
\end{abstract}
\begin{IEEEkeywords}
Remediation, Compatibility, Java, Open-source software
\end{IEEEkeywords}

\section{Introduction}
\label{introduction}

The exposure of open-source third-party libraries (TPLs) vulnerabilities in recent years, such as the well-known Log4Shell vulnerability \cite{log4jvul,log4jvulrce}, has been drawing increasing attention. To accurately detect the versioned TPLs and the disclosed vulnerabilities in users' projects, software composition analysis (SCA) \cite{sca} has been widely applied to scan projects and return detected TPLs for security analysis. The detection has been well developed and implemented in various academic and commercial SCA tools \cite{steady,dependabot,snyk,whitesource,owasp,scantist}. However, the remediation to fix vulnerabilities in TPLs by version adjustments has no broadly acknowledged solution but various strategies. 

We further investigated existing tools. Community tools, 
such as \dpb \cite{dependabot}, only considers vulnerabilities of direct dependencies. Other popular anonymous commercial tools use reachability analysis as the prioritization metric, but none considers the compatibility of upgrades of the dependencies. An academic tool, \steady, calculates the percentage of changed classes or methods as the probability for compatibility, which is inaccurate by its nature of uncertainty. 

Due to different strategies, the effectiveness of remediation tools varies substantially, which will be clarified in the preliminary study and the evaluation. Moreover, the side effects of remediation could hinder the adoption of suggestions by users. 
According to our study \cite{dataset} of the rejected remediation suggestions at GitHub,
the primary concern of users was incompatibility, which accounted for $51.31\%$.

Unfortunately, these concerns, especially compatibility, cannot be appropriately handled by existing remediation tools due to two reasons: (1) They conduct local optimization on individual libraries instead of the global optimization of the entire dependency graph (\depg), which may miss incompatible relationships and fail to handle the trade-off between compatibility and security. (2) They offer suggestions based on the original \depg and overlook the structural changes that suggestions bring to it and the underlying call graphs. As a result, the outdated \depg could lead to incompatibility, lack of remediation on new vulnerabilities, and wasted remediation on unused dependencies.

To address the problems of existing tools and achieve remediation of better quality, three major \textbf{challenges} have to be resolved: 
\textbf{c1}: The absolutely optimal solutions for libraries are not always available. So the trade-off between security and compatibility during decision-making has to be handled. 
\textbf{c2}: The complexity of global optimization increases exponentially with the number of dependencies because the version combinations over all libraries should be traversed. 
\textbf{c3}: The suggestions on one library can, directly and indirectly, change the \depg structure, call graphs, and compatibility of \depg. Accordingly, the optimal solutions for the rest of the libraries may also be changed. These effects propagate from the changed library to the entire \depg through dependency relationships, referred to as \rip in this paper. 
The \rip may lead to sub-optimal solutions if DG is not updated accordingly.

To tackle the above-mentioned challenges, we propose \underline{Co}mpatible \underline{R}emediation of Third-p\underline{a}rty \underline{L}ibraries (\tool) to remediate vulnerabilities in dependencies by version suggestions without breaking the projects  with a balanced time cost for Maven \cite{mvn}. 
\tool starts with the \depg and the underlying call graphs of the target project. Then, \tool splits the \depg into subgraphs with two steps of partitioning for \textbf{c1}. \tool walkthroughs subgraphs with a top-down approach and calculates the best solutions with SMT solver within subgraphs for \textbf{c2}. During the walkthrough, subsequent subgraphs are dynamically updated for the \rip to handle \textbf{c3}. To avoid dead ends, backtracking mechanisms are implemented in \tool.

We have evaluated \tool by comparing it with
state-of-the-art remediation tools regarding security and compatibility. It turned out \tool fixed the most vulnerabilities ($87.56\%$) among all tools (best of others $75.32\%$) and achieved the best assurance of compatibility ($98.67\%$ successful compilation rate and $92.96\%$ successful unit test rate).
Moreover, the designs of subgraph partitioning and the trade-off between compatibility and security were evaluated against the baselines.
The result showed \tool broke fewer projects and spent much less time than them at the cost of $4.05\%$ fewer vulnerabilities fixed. Furthermore, we found that $78.45\%$ of vulnerabilities in popular Maven projects could be fixed without breaking the projects. However, without the aid of \tool, only $25.71\%$ could be straightforwardly fixed by users. The contributions we have made are as follows.

\begin{itemize}[leftmargin=9pt]
    \item We proposed \tool as a remediation tool for Maven projects to handle the global optimization for enhanced security and compatibility.
    \item We studied the concerns of users regarding remediation suggestions by analyzing Pull Requests (PRs) and found that $51.31\%$ of cases were related to incompatibility. 
    \item We empirically compared 
    and analyzed strategies of popular remediation tools regarding their support of compatibility and prioritization for the reference of other researchers.
\end{itemize}

\section{Motivations}

\subsection{Motivating Example}
\label{sec:background}
\dpb has been widely used as the most popular dependency security management extension at GitHub. One of the most popular Maven projects, \textit{commons-lang} \cite{commonlang}, adopted \dpb to manage their dependencies. Nevertheless, the remediation caused build failure after upgrading \cite{rembuildissue}.
\dpb has implemented the compatibility score by calculating the test passing rates from other repositories as the confidence score. However, in this case, the compatibility score was \textit{unknown}. The compatibility score relying on knowledge of the crowd cannot guarantee a successful compilation without code-based compatibility calculation. Thus, \tool relies on static code-based compatibility checkers aligning with a global perspective of \depg to ensure the adjusted dependencies do not break the project.
Motivated by the motivating example, we studied the strategies of state-of-the-art remediation tools to understand how existing tools handle incompatibility issues. Then, we further studied the concerns of users regarding the remediation suggestions at GitHub to recognize what can be improved.

\subsection{Study of Remediation Strategies of Existing Tools}
To understand the implicit reasons for the breaking in \Cref{sec:background}, we first empirically compared the published remediation strategies of existing tools and then quantitatively evaluated them in \Cref{sec:evaluation}. We only counted tools that provided actionable advice for dependencies, while tools that only offered multiple suggestions for vulnerabilities were out of the discussion because users would have to select the version out of multiple suggestions manually during decision-making for each library. The tools included \dpb, \steady, and two popular commercial tools denoted by \textit{\snyk} and \textit{\scantist}.

\begin{table*}
\scriptsize
\footnotesize
\caption{Comparison of State-of-the-art SCA Tools That Provide Remediation}
\setlength{\tabcolsep}{6pt}
\begin{tabular}{lllllllll}
    \toprule
\textbf{Tool}  & \textbf{Fix level} & \textbf{Fix target} & \textbf{Compatibility} & \textbf{S \& C trade-off} & \textbf{Reachability} & \textbf{Dep conflict} & \textbf{Ripple effects} & \textbf{Unused dependencies}\\
    \midrule
\steady     & All graph   & Vertex      & \halfcirc   &    Sec first   & \fullcirc            & \emptycirc                  & \emptycirc  & \emptycirc   \\

\dpb & Direct    & Vertex      & \halfcirc &    Sec first     & \emptycirc            & \emptycirc            & \emptycirc    & \emptycirc  \\
\snyk       & Direct   & Vertex      & \emptycirc     & Sec only        & \fullcirc            & \emptycirc             & \emptycirc  & \emptycirc  \\
\scantist  & Direct   & Tree       & \emptycirc & Sec only         & \emptycirc            & \emptycirc        & \halfcirc & \emptycirc  \\
		\bottomrule
\end{tabular}
\label{tab:comparisontool}
\begin{enumerate}
    \item \textit{Fix level}: direct/direct+transitive dependencies. \textit{Fix target}: basic units that the tools consider during optimization. \textit{S\&C trade-off}: prioritization of security or compatibility during version determination. \textit{Ripple effects}: the support to handle the side effects brought by \rip.
\end{enumerate}
\end{table*}

\begin{itemize}[leftmargin=9pt]
\item \textbf{\dpb}: \dpb is able to create PRs to upgrade vulnerable dependencies to clean versions instead of providing an overall suggestion for the entire \depg.
As for the compatibility, \dpb calculates the successful test rate of the upgrades from other repositories as the confidence score. However, this score can be unreliable because it is usually unavailable, and the compatibility ultimately depends on the context of the code base.
\item \textbf{\steady}: \steady is an open-source academic SCA tool with an open-source vulnerability database. \steady adjusts the versions of both direct and transitive dependencies to reduce the vulnerability risks at a fine granularity. Also, it utilizes the reachability analysis of vulnerabilities to filter out the unreachable CVEs with low risks. The reachability comprises both static and dynamic analysis, which only constructs call graphs once at the beginning.
As for the version selection, \steady prioritizes the non-vulnerable versions, then determines the best candidate with the compatibility probability $p$. To derive $p$, it defines the reachable constructs (class, method, etc.) as \textit{touch points} and calculates the percentage of present \textit{touch points} in upgraded versions as $p$. The probability could be unreliable due to its uncertainty.
\item \textbf{\snyk}: Towards a \depg, \snyk tweaks only the direct dependencies to remediate the vulnerabilities. The fundamental strategy is to upgrade the libraries with vulnerabilities to the closest non-vulnerable versions, as the closer versions usually are more likely to be compatible. The reachability is implemented by WALA \cite{wala} in a static manner to prioritize the critical reachable vulnerabilities. However, the compatibility of the remediation is not taken into account.
\item \textbf{\scantist}: \scantist conducts the remediation on the direct dependencies. The key feature is that \scantist considers all vulnerabilities of transitive dependencies associated with the direct dependencies. Specifically, it iterates over all direct dependencies. For each, \scantist attempts the version candidates and resolves the subsequent dependencies to measure the updated overall vulnerabilities. Then, \scantist selects the version with the fewest overall vulnerabilities for this direct dependency. The strategy considers the \rip from the upgraded direct dependency to the upstream tree. However, as direct dependencies are usually not independent but inter-connected by transitive dependency relationships, the respective optimization of each direct dependency does not necessarily result in global optimization. 
\end{itemize}

The comparison of SCA tools is demonstrated in Table~\ref{tab:comparisontool}. \textit{Fix level} refers to the direct/transitive dependencies to be fixed. \textit{Fix unit} denotes the basic units that the tools optimize. \textit{S\&C trade-off} means the prioritization of determining the best candidates. The rest of the columns are summarized in the next section. 
From the remediation strategies of tools, we found three major causes of incompatibility issues. \textbf{(1)} \textbf{No reliable detection}: Although \steady and \dpb support compatibility scores, their results were unreliable due to inaccuracy. \textbf{(2)} \textbf{Lack of global optimization}: Because vertices in \depg were interconnected with each other, optimizations of them were not independent. Thus, it is impractical to optimize each vertex individually without a global perspective.
\textbf{(3)} \textbf{Lack of support of handling \rip}: The optimization was conducted based on the original \depg without updating structures and call graphs. Then, the optimal solutions based on the new \depg were changed so that the existing tools would return sub-optimal solutions.

\subsection{Study of Users' Concerns with Remediation Suggestions}
\label{sec:mavenstudy}
GitHub provides various automated SCA extensions to create PRs of security updates for dependencies, but these PRs are far from perfect, and thus sometimes rejected by users. 
To increase the acceptance rate of suggestions, we conducted a study to understand the concerns of users towards the remediation at GitHub by analyzing the reasons for rejected remediation suggestions and the accepted suggestions as a comparison. 

Due to the lack of existing studies on Maven projects, the data set was collected by ourselves. First, we derived $9,527$ projects active in the last three years with 100+ starts at GitHub. Then, $5,356$ un-merged PRs created by bots were located and narrowed down to $306$ PRs with human participation. Finally, we manually went through the comments in these PRs and summarized several reasons why PRs were unmerged.
\begin{itemize}[leftmargin=9pt]
    \item ($91$ cases, $29.74\%$) \textbf{Duplication}: The upgrades were superseded by other PRs, which were eventually merged.
    \item($82$ cases, $26.80\%$) \textbf{Compilation/Test/CI failures and Dependency conflict (DC)}: The developers ran tests on the projects with upgraded dependencies, and incompatible issues occurred. Particularly, tests failed at dependency resolution, compilation, and test stages. For all PRs created by \dpb in this category, compatibility scores were shown as \textit{unknown}. 
    \item ($75$ cases, $24.51\%$) \textbf{Incompatibility concerns}: The developers were concerned by incompatibility risks because either the upgrades had large spans, such as major upgrades, or they were known to be breaking. All compatibility scores were shown as \textit{unknown} as well.
    \item (23 cases, $7.51\%$) \textbf{Internal errors}: Bots reported their internal errors in comments, so the users closed the PRs.
    \item (12 cases, $3.92\%$) \textbf{Unused dependencies}: The developers found the dependencies to be upgraded were not in use anymore, so the PRs were closed. The \textit{bloated dependencies} were supposed to be ignored during the remediation. 
    \item (9 cases, $2.94\%$) \textbf{Disobeying rules or absence of signed agreements}: The developers closed the PRs because the PRs failed to follow the rules of the repositories or sign the contributor agreements.
    \item (8 cases, $2.61\%$) \textbf{Unknown reasons}: The developers closed the PRs without explicitly mentioning the reasons.
    \item (6 cases, $1.96\%$) \textbf{Other}: There were various reasons: (1) Upstream projects demanded to keep the current version. (2) Java version was not compatible. (3) The PR introduced new CVEs. (4) Wrong user configuration. (5) A formatting issue. 
\end{itemize}
From the result, excluding the duplicated PRs and unrelated reasons, such as internal errors, it is evident that the compilation/test failures and incompatibility concerns were the primary concerns of users ($51.53\%$). The upgrades on unused dependencies could be avoided by the reachability analysis. The perspectives of concerns of users are demonstrated in Table~\ref{tab:comparisontool}. \textit{Dep conflict} refers to the support of the detection of possible dependency conflicts raised by Maven. The \rip denotes the support of dynamically handling the \rip. \textit{Unused dependencies} means the support of detecting and ignoring unused dependencies.

Besides the reasons for rejected PRs, merged PRs were also studied as a comparison, but they usually failed to include the reasons for acceptance. Thus, we studied the distribution of their upgrades. Since the number of merged PRs is enormous, we studied the $556,257$ PRs merged in the last two years for Maven projects. The distribution was (1) \ma: $11.91\%$; (2) \mi: $38.34\%$; (3) \pa: $48.55\%$; (4) \textit{pre-release}: $0.89\%$; (5) No SemVer available: $0.31\%$. The result indicated that most merged PRs ($87.79\%$) did not bump the versions to major upgrades, which followed the criteria of SemVer because non-major upgrades were supposed to maintain backward compatibility. Therefore, the remediation suggestions with fewer major upgrades are more likely to be accepted by users.

\section{Methodology}

\subsection{Problem Formulation}
By summarizing the users' concerns, we are able to define the objectives and constraints of the remediation. The primary objective is to minimize the total vulnerability risks:
\begin{equation}
\label{eq:vul}
min\quad F_{vul} = \sum_{m=1}^M \sum_{vul=1}^{Vul} \theta_{vul} f_{cvss}(vul) 
\end{equation}
where $M$ is the number of libraries 
and $Vul$ is the number of vulnerabilities of a vertex m. $f_{cvss}$ is the Common Vulnerability Scoring System (CVSS) 
\cite{cvss} weight. $\theta_v$ is the reachability coefficient for vulnerability $v$, particularly, $\theta_v$ is larger for reachable vulnerabilities,
because the reachable vulnerabilities are possible to be exploited by attackers. 
However, in reality, not all vulnerabilities are open-source, which also increases the difficulty for attackers. Thus, the vulnerabilities with uncertain vulnerable classes or methods are classified as \textit{unknown vulnerabilities} whose severity is ranked between the reachable and unreachable vulnerabilities. 
Since different vulnerabilities result in different risks, we use CVSS, a normalized score provided by 
NVD, to prioritize the vulnerabilities with higher risks during calculation.

The remediation is less likely to be accepted if it breaks the users' projects, according to the study in \Cref{sec:mavenstudy}. Thus, the pre-condition of successful remediation is the compatibility of version adjustments. 
\begin{equation}
\label{eq:incom}
    s.t. \quad c_{incom} = \sum_{m=1}^M  \sum_{p=1}^P \theta_{v} * incom(v_p, v_m) = 0
\end{equation}
$P$ is the number of parent vertices of $v_m$, while $v_p$ is a parent vertex. $c_{incom}$ is the total number of dependency relationships that cause incompatible issues. The incompatibility comprises two types of code-based breaking (semantic and syntactic breaking) and DC issues.

To achieve the global optimization and handle the \rip mentioned above, \tool is supposed to optimize all connected vertices altogether in a dynamically adjusted \depg. These goals bring three challenges: (1) Trade-off between the security and the compatibility during decision making. (2) The time complexity increase exponentially with the size of \depg as $O(n) = \prod_{n=1}^{N}$ if all solutions are to be iterated over. (3) The \rip requires dynamically updated \depg.

\begin{figure}[!t]
\centering
  \includegraphics[width=0.9\linewidth]{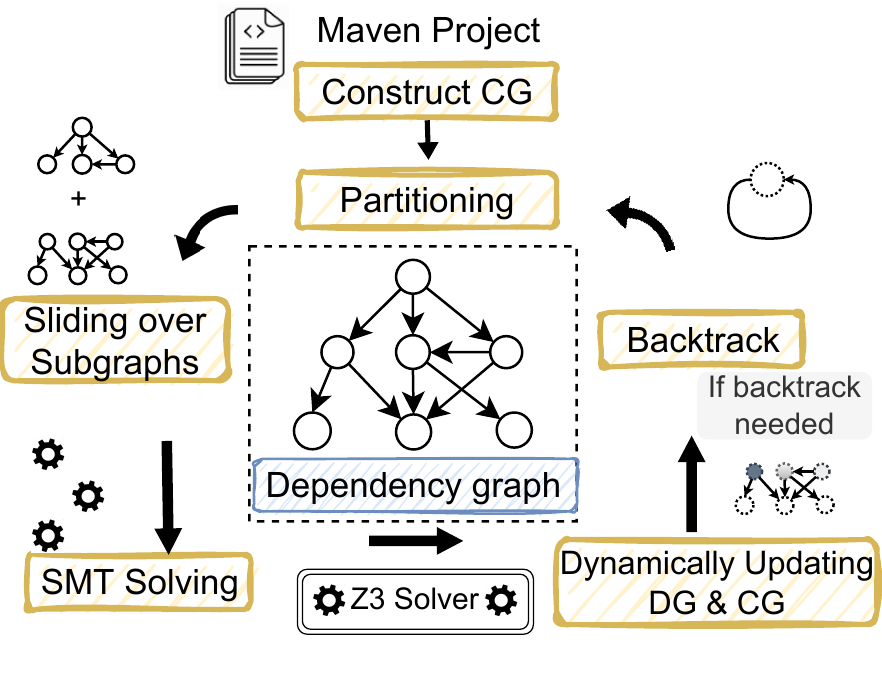}
  \caption{Overview of \tool} 
  \label{fig:overview}
\end{figure}

\subsection{Overview} 
\label{sec:overview}
\tool is implemented in four steps as illustrated in Fig.~\ref{fig:overview}. (1) Generating \depg and the call graph (CG) from the project object model (pom) file, a version control file of Maven, and class files of the project. (2) Partitioning the \depg into subgraphs. (3) Optimizing the subgraphs regarding the vulnerability risks based on the pre-computed vulnerability mappings while ensuring compatibility. (4) Backtracking to parent vertices heuristically if the dead end is met. Then, the final remediation suggestions of version adjustment of all TPLs in the \depg are returned.

\subsection{Constructing Dependency Graph and Call Graph}
With pom files and class files, \tool extracts the dependency tree by the Maven command and recovers the \depg by completing the absent dependency relationships from a pre-computed dependency database. According to Maven documentation \cite{mvnscope}, as dependencies with \textit{test} scope are not involved in the normal use of the projects, \tool excludes dependencies with \textit{test} scope from the \depg.  Specifically, \depg is represented as $DG = Graph(V, E)$, where $V=\{e_i^x \mid i\in\{0,...,N-1\}, x\in\{0,...,L\}\}$ and $E=\{e_i \rightarrow e_j \mid i,j\in\{0,...,N-1\}\}$.
$\rightarrow$ denotes the direction of the calling edge, and $x$ specifies the stack level w.r.t the \depg.

The CG is constructed statically based on Soot \cite{vallee2010soot} by the Spark algorithm \cite{spark} from the class files of the projects. The \textit{main} methods are considered the entry points which serve as the start of the call graphs. If \textit{main} methods are absent, we overestimate that it is possible to execute all methods implemented in the projects. Thus, all methods in users' projects are considered entry points. Since handling the \rip requires the dynamically updated CG to achieve real-time reachability analysis, the call edges in the CG are collected modularly. i.e. call edges are not extracted from a Uber jar \cite{uberjar} (root project with all dependencies) but from jars of each dependency separately and sequentially and then integrated into one graph originating from the root project. Particularly, for each dependency, the callers from the parent libraries serve as the entry points for child libraries. After the remediation, if the child libraries are suggested for other versions, the callees in them can be substituted accordingly to generate the real-time CG flexibly.

\begin{figure}[!t]
  \includegraphics[width=1\linewidth]{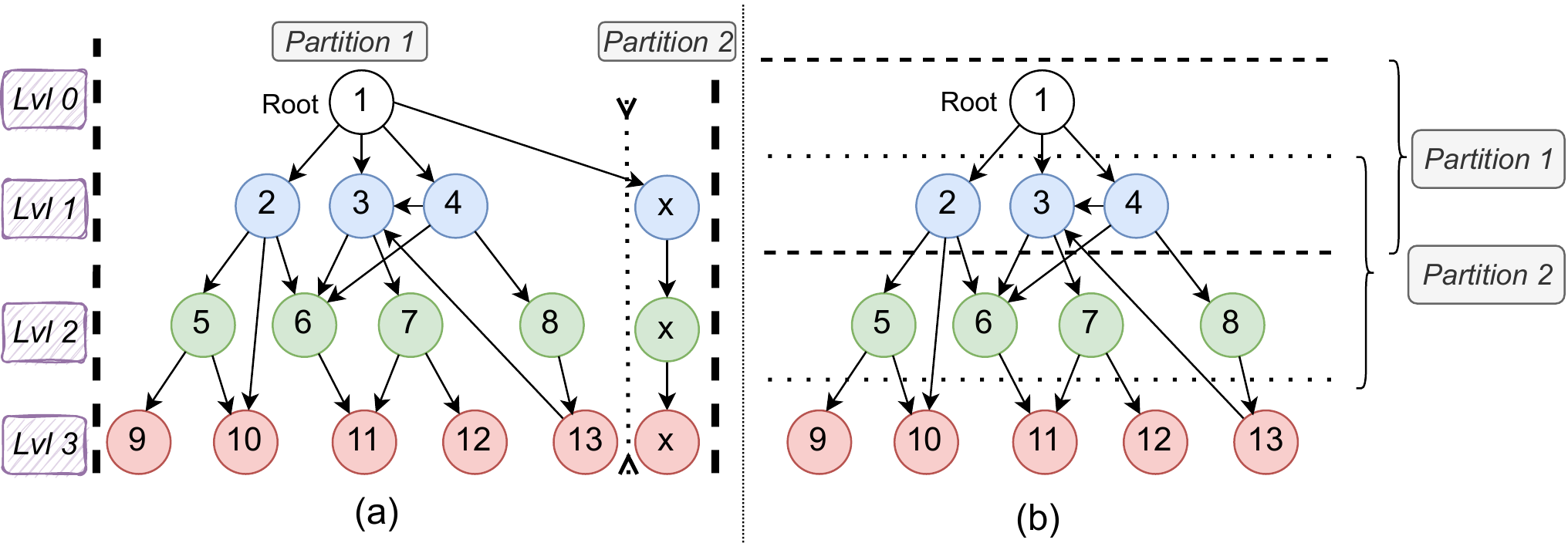}
  \caption{Dependency Graph Vertical (a) and Horizontal (b) Partitioning} 
  \label{fig:partition}
\end{figure}

\subsection{Partitioning Dependency Graph}
Due to the high complexity of optimization over the entire \depg, \tool partitions the \depg into subgraphs to reduce the size of the overall solution space. The partitioning comprises two steps, vertical partitioning, and horizontal partitioning. As illustrated in Fig.~\ref{fig:partition} (a), the vertical partitioning iteratively splits the \depg into multiple partitions that are not connected with each other by dependency edges except the direct relationships from the root project $v_1$ {until all unconnected partitions are split. }Since the direct dependencies do not depend on each other, optimizations on multiple partitions can be conducted independently and concurrently. For example, in Fig.~\ref{fig:partition} (a), \textit{partition 1} and \textit{partition 2} do not depend on each other. Hence, they can be partitioned to boost performance.

However, the vertical partitioning is not always sufficient, especially for the large partition at left in Fig.~\ref{fig:partition} (a). In this case, horizontal partitioning can further reduce the solution space. The subgraphs are partitioned by levels to preserve the semantics. According to \cite{schroter2010stack}, the semantics of a method decays along the calling chain, i.e. dependencies closer to the root matter more than those farther from the root in terms of the semantics or functioning they provide. For better notations, dependencies are labeled by tags called \textit{level} to denote the smallest number of hops from the root. To better preserve the semantics of dependencies against the potential incompatibility, \tool split \depg and group vertices at level $l$ and $l-1$ into subgraphs as in Fig.~\ref{fig:partition} (b). Then, because the closer dependencies preserve more semantics, \tool starts the optimization from the root user projects in a top-down manner. 
Particularly, the lower-level dependencies should humor the upper ones in terms of the compatibility constraints as much as possible. 
Hence, \tool attempts to optimize dependencies in two adjacent levels at a time and then moves the sliding window of a partition down to the next level with a newly updated CG. With the horizontal partition, the complexity can be reduced to $O(n) = \sum_{p_{hori}=1}^{P_{hori}} \sum_{p_{vert}=1}^{P_{vert}} \prod_{n=1}^{N_p}$. The side effect is that the potential better solution with lower vulnerability risks may be overlooked for dependency edges across multiple levels. To compensate for the loss, \Cref{sec:backtrack} introduces the backtracking mechanisms to avoid sub-optimal situations. 

\subsection{Optimizing Subgraphs}
In this subsection, the detailed specification of the optimization on subgraphs based on Z3 SMT solver \cite{z3solver} is described. 
\subsubsection{\textbf{Objectives and Constraints Definition}}
In each subgraph, \tool conducts the optimization to minimize the vulnerability risks in the condition that the version changes are compatible. The vulnerability elimination follows the objective function in \Cref{eq:vul}. The basic vulnerability elimination strategy is to find versions with the fewest reachable and unknown vulnerabilities. Then, if more than one versions satisfy these conditions and other constraints, the versions without unreachable vulnerabilities are preferred. 

\begin{figure}[!t]
  \includegraphics[width=1\linewidth]{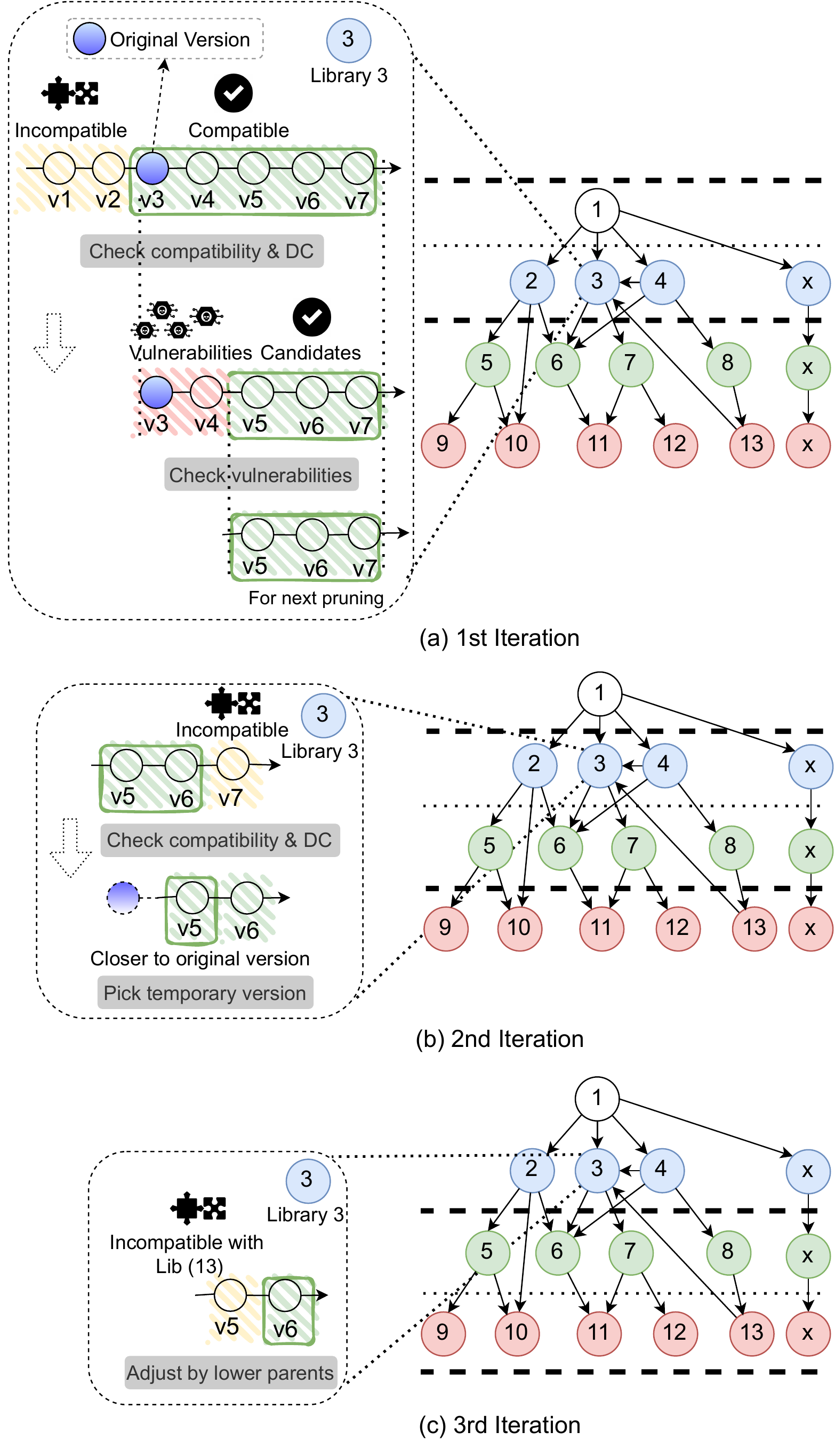}
  \caption{Example of the Version Selection of a Dependency Library} 
  \label{fig:verselect}
\end{figure}

Theoretically, the compatibility constraint is supposed to be strict. 
However, not all types of incompatibility can be accurately detected. Generally, there are three major types that \tool aims to resolve, namely, semantic breaking, syntactic breaking, and dependency conflicts, as discussed in \Cref{sec:mavenstudy}. Except for semantic breaking, the rest can be detected statically and efficiently. Thus, the detection of the rest is integrated into the optimization as constraints:
\abovedisplayskip=1pt
\abovedisplayshortskip=1pt
\belowdisplayskip=1pt
\belowdisplayshortskip=1pt
\begin{align}
\label{eq:synb}
    s.t. \quad c_{synb} = \sum_{m=1}^M \sigma * synb(P(x'_m), x'_m) = 0
\end{align}

The $synb$, Syntactic Breaking, is calculated based on the reachability analysis and the API compatibility checkers. For each version pair of one library, the modified APIs that can cause the failure of compilation are calculated by the three most widely used API compatibility checkers japi-compliance-checker \cite{japi-compliance-checker}, revapi \cite{revapi}, japicmp \cite{japicmp} based on the pair of jar files. Then, based on the reachability analysis, the called APIs of this library are obtained from CG. If any problematic APIs are called, the compilation would mostly fail, so \tool would label this candidate version as breaking and discard it.
\abovedisplayskip=1pt
\abovedisplayshortskip=1pt
\belowdisplayskip=1pt
\belowdisplayshortskip=1pt
\begin{align}
\label{eq:dc}
    s.t. \quad c_{dc} = \sum_{m=1}^M dc(P(x'_m), x'_m) = 0
\end{align}

The DC issues are calculated based on Maven version rules \cite{mavenver}. Like other package managers, version ranges define the allowed versions for dependencies. 
If two version ranges required by dependents do not overlap, Maven would report \textit{Dependency Conflict} during version resolution before the compilation. A similar logic is implemented in \tool to only select versions within the intersection of ranges defined by dependents. It is noteworthy that over $99\%$ dependency version specifications are not determined with ranges, but single recommended versions instead, which means all versions are available regardless of compatibility. In this case, \tool would include all versions as candidates for DC detection as Maven.

Since the semantic breaking is usually revealed by unit tests subject to limited coverage according to \cite{chen2020taming}, it is hard to detect it statically and efficiently. Also, it is the leading cause of unit test failures \cite{mostafa2017experience}, which is one of the main reasons why users reject remediation suggestions. Thus, \tool relies on auxiliary information to infer the potential semantic breaking and minimize its probability by following the SemVer and Maven versioning guides. According to SemVer, the \ma upgrades are allowed to break the original implementations. Hence, \tool avoids using \ma upgrades/downgrades as much as possible unless they are less vulnerable and satisfy the other compatibility criteria. Thus, besides the primary objective, we add a secondary objective, $f_{major}$, the number of dependencies that have \ma upgrades/downgrades:
\begin{align}
\label{eq:maj}
       min \quad f_{major} &= \sum_{n=1}^N f_{major,x_n}(x_n, x'_n)\\ \text{ where }f_{major} &= \begin{cases}
      0 & \text{if $x_n$ to $x'_n$ is not major}  \\
      1 & \text{if $x_n$ to $x'_n$ is major} 
    \end{cases}  \nonumber
\end{align}
{Although SemVer stipulates \mi should not include incompatible changes, researchers from \cite{xavier2017historical} found that \mi upgrades are not as compatible as \pa upgrades, which generally introduce more breaking changes. 
Therefore, \tool always prefers \pa upgrades rather than \mi if all other conditions stand. Another secondary objective function of $f_{minor}$ is created to fulfill the purpose.}
\begin{align}
\label{eq:min}
       min \quad f_{minor} &= \sum_{n=1}^N f_{minor,x_n}(x_n, x'_n)\\ \text{ where }f_{minor} &= \begin{cases}
      0 & \text{if $x_n$ to $x'_n$ is not minor}  \\
      1 & \text{if $x_n$ to $x'_n$ is minor} 
    \end{cases}  \nonumber
\end{align}
Besides SemVer, Maven version control rules \cite{mavenver} also help identify potentially breaking versions. First, the pre-release versions, also known as development versions, such as \textit{alpha, beta, SNAPSHOT} versions, are unstable and prone to breaking changes, which are selected at a lower priority than \ma upgrades. Second, the larger version spans are usually more likely to induce incompatible changes. \tool attempts to reduce the version span from the original version to the new version as much as possible. In terms of these two objectives, the functions of $f_{dev}$ and $f_{span}$ are formally given as:
\begin{flalign}
\label{eq:dev}
       min \quad f_{dev} &= \sum_{n=1}^N f_{dev,x_n
}(x_n, x'_n)
       \\  \text{ where }f_{dev} &= \begin{cases}
      0 & \text{if $x'_n$ is not dev }  \\
      1 & \text{if $x_n$ is not dev, $x'_n$ is dev} \nonumber
    \end{cases}  \nonumber
\end{flalign}

\begin{flalign}
\label{eq:span}
       min \quad f_{span} &= \sum_{n=1}^N dist(x_n, x'_n)
\end{flalign}
where $dist(x, y)$ is the distance between x,y in sorted versions.

After solving with the SMT solver, each vertex in the subgraph is assigned with a selected version, and upgraded libraries in CG will be updated accordingly. However, the selected versions can be overthrown by the next optimization. Thus, all selectable candidate versions are saved and fed to the next optimization. For instance, in Fig.~\ref{fig:verselect} (a), \textit{Lib $3$} initially has $7$ candidates and gets filtered to $3$ by incompatibility and vulnerabilities. In the next iteration (b), \textit{Lib $3$} has its candidates further filtered to $2$ because of the incompatibility. Then, $v5$ is selected due to its smaller version span from the original version. However, in the third iteration (c), $v5$ is overthrown because it is not compatible with the parent library \textit{Lib 13} at a lower level. Since the compilation and Maven resolution would fail regardless of the levels, the selected versions must follow the constraints in \Cref{eq:synb,eq:dc}. Therefore, $v5$ is discarded, and $v6$ with compatible changes is selected.

\subsection{Backtracking}
\label{sec:backtrack}
Although sequential partitions of \depg reduce the complexity, they could lead to sub-optimal solutions and dead ends. To mitigate such issues, two types of backtracking mechanisms are implemented in \tool, the hard and the soft backtracking. 

\subsubsection{\textbf{Hard Backtracking}}
Hard backtracking is implemented to avoid dead ends. It happens during deciding the best version of a library where all versions disobey the constraints by potentially breaking the project. The backtrack targets are parent libraries of the current library. Since backtracking requires re-visiting the related vertices, the parent library at the lowest level is prioritized to reduce the efforts of re-visiting. And then, the higher ones are attempted if the lower parent triggers the backtrack again. During one backtrack, the selected version of the target parents is temporarily marked as incompatible, and other versions are attempted.
\begin{algorithm2e}[t!]
\setstretch{0.9}
    \footnotesize
 \setcounter{AlgoLine}{0}
 \caption{Algorithm of \tool}
 \label{alg:rem}
 \DontPrintSemicolon
 \SetCommentSty{mycommfont}
 {
     \KwIn{Dependency Graph $G \left \langle V,E \right \rangle$ (vertices $V$ and edges $E$) with $h$ levels, class files $cf$ of the project}
     \KwOut{Remediated $G' \left \langle V',E' \right \rangle$ with newly assigned versions}
     $parts_v \gets verticalPartition(G)$\;
     
     \ForEach{$part$ in $parts_v$}{
         \ForEach{$i_{th}$ in $h$}{
            $part_h \gets V_i + V_{i+1}$\;
            $cg \gets CallGraph(part_h, cf)$\;
            
            \ForEach{$v$ in $V$}{
                $parents \gets parentsOf(v)$\;
                
                \ForEach{$ver$ in $versionsOf(v)$}{
                    \If{ver has synb \textbf{or} DC}{
                        $cand.remove(ver)$\;
                    }
                    \If{$sizeOf(cand)==0$}{
                        hardBacktrack\;
                        break \;
                    }
                    $vuls \gets vulsOf(ver)$ \;
                    
                    \ForEach{$vul$ in $vuls$}{
                        $\theta \gets reachability(vul, cf)$\;
                    }
                    sort candidates by $\theta$ \;
                }
                $s \gets SMTsolver(V_i, V_{i+1})$\;
                \If{$vuls(s) != min(vuls))$}{
                    softBacktrack \;
                    break \;
                }
                $cg \gets updateBy(s)$\;
                
                $G \gets updateBy(s)$\;
                
                \If{hardBacktrack}{
                    $p \gets  parent_{lowest}$ \;
                    
                    $p.incompatible \gets ver$\;
                    
                    backtrack to p \;
                }
                \If{softBacktrack}{
                    $p \gets parent_{lowest}$  \;
                    $runs \gets saveVul(p)$\;
                    backtrack to p \;
                    
                    \ForEach{$r_{th}$ run in $p.vers$}{
                        $ runs \gets saveVul(p_r)$\;
                    }
                    $s \gets min(runs)$\;
                }
            }
        }
    }
    \KwRet{$G' \left \langle V',E' \right \rangle$}
}
\end{algorithm2e}

\subsubsection{\textbf{Soft Backtracking}}
Soft backtracking is used to avoid sub-optimal solutions. It is triggered when the version selected by the SMT solver is not the version with the lowest vulnerability risks in the version list, such as non-vulnerable versions. Like the hard backtrack, the soft backtrack prioritizes the parent libraries at lower levels. The different part is that soft backtrack does not mark the parent's current version as incompatible but unpreferrable instead. It means if other versions are proven to be not as optimal as the unpreferrable version after the backtracking, the unpreferrable would still be selected. Thus, even if versions satisfy the constraints, they could be ignored by soft backtracking. During the soft backtracking, \tool saves the overall vulnerabilities between the backtracked library and the target parent for future comparison. After the backtracking, \tool compares the vulnerabilities of the current run with the ones saved previously and adopts the run with the fewest vulnerabilities to apply the versions to backtracked libraries accordingly. Note that to avoid an infinite loop, soft bakctracking would not be triggered again during one run of soft backtracking. Also, if the hard backtracking is triggered during soft backtracking, the current run would be discarded, and other versions would be attempted.

In conclusion, \tool was designed to overcome the challenges of the high complexity of global optimization and \rip. The algorithm is presented in \Cref{alg:rem}. \tool starts with vertical and horizontal partitions to split the \depg into multiple parts. Then, the SMT solver is used to optimize the remediation results in each partition in a top-down manner. If any backtrack is triggered, \tool backtracks to the previous vertices to avoid the sub-optimal solutions.

\section{Evaluation}
\label{sec:evaluation}
We aim to answer the following research questions:

\noindent \textbf{RQ1}: How is \tool compared with other cutting-edge remediation tools regarding security and compatibility?

\noindent \textbf{RQ2}: How effectively does \tool resolve the challenge of global optimization by subgraph partitioning?

\noindent \textbf{RQ3}: How many vulnerabilities CAN/CANNOT be fixed without breaking the projects in the Maven ecosystem?

\subsection{Preparation}
\subsubsection{\textbf{Data Collection}}
\begin{figure}[!t]
  \centering \includegraphics[width=0.9\linewidth]{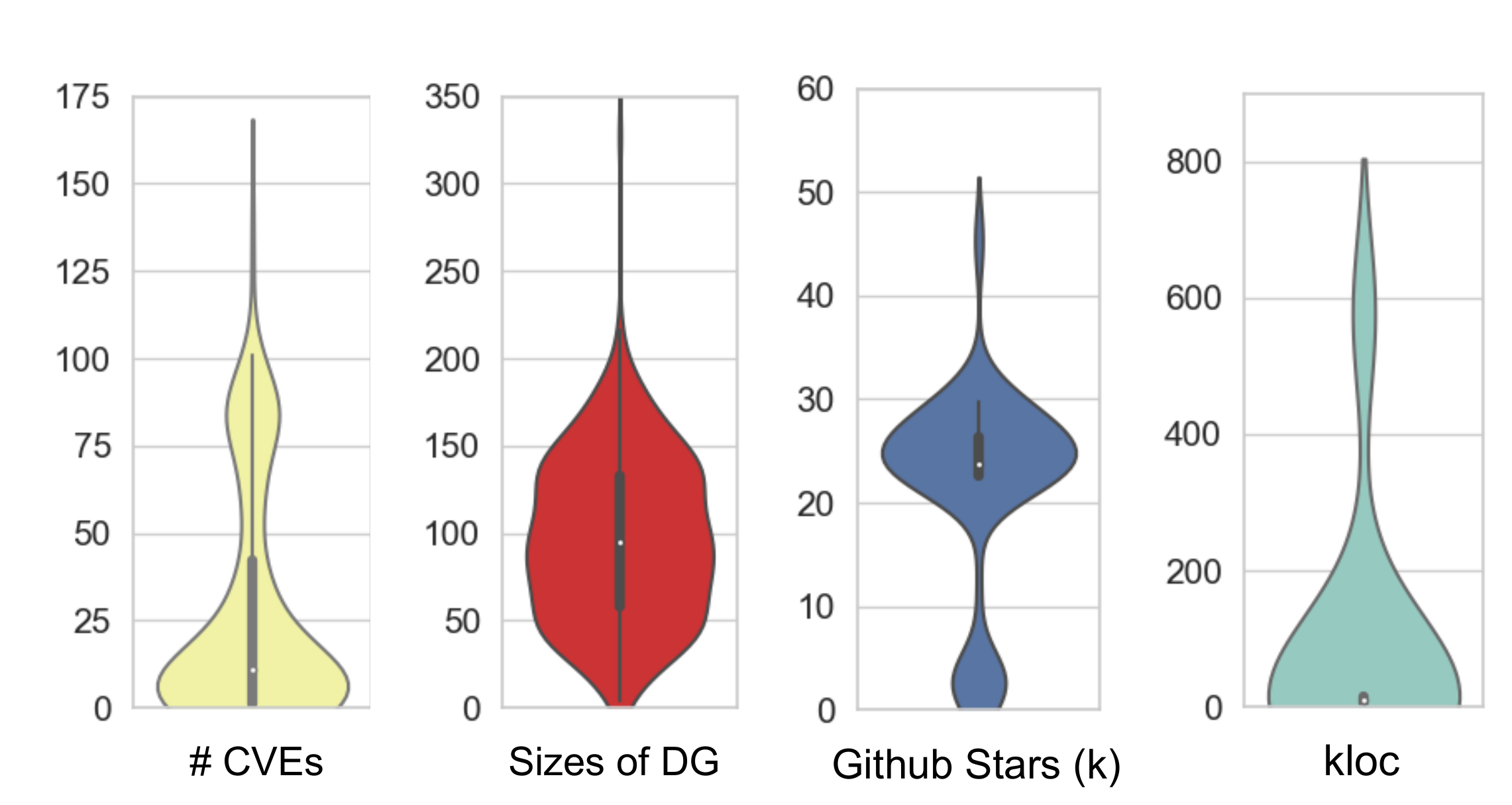}
  \caption{Demographics of the Data Set of RQ1 and RQ2} 
  \label{fig:demo}
\end{figure}

To build a data set of in-development Maven projects, we collected $301$ most starred projects managed by Maven at GitHub on May 21st, 2022. {We first selected Java projects with the most stars from GitHub and excluded non-Maven projects. Next, we manually modified the POM files of each project to apply the remediation suggestions from these tools. Considering the efforts of manual work, we filtered these projects with 1K+ stars. Finally, we got 301 selected projects.} The demographics of the data set are illustrated in Fig.~\ref{fig:demo}. It has the following features: (1) The code base size is non-trivial (average $22.19$ kloc). (2) The range of sizes of dependency graphs is large (max $327$, average $32.0$). (3) The projects are affected by an adequate number of CVEs (average $27.6$). (4) The projects are popular because they are highly starred. 

To experiment with accurate vulnerability mappings, we periodically crawled CVE feeds from NVD \cite{nvd} with a pipeline and pre-classified the language of CVEs by keyword matching. As the CVE descriptions are free-text~\cite{guo2022detecting,guo2021key}, it is impractical to directly extract version mappings from them. Hence, we manually triaged the mappings from reference links and associated Common Platform Enumerations (CPEs) \cite{cpe}. So far on May 21st, 2022, we collected mappings for $1,759$ CVEs associated with Maven libraries. In this section, the evaluation needs the reachability analysis, which requires the vulnerable methods and classes associated with CVEs. Thus, we first identified $750$ CVEs ($42.64\%$ of all Maven CVEs) from $2,326$ unique libraries used as dependencies in $301$ projects. Then, vulnerable classes and methods of $300$ CVEs were successfully identified and manually collected from the patches available at NVD links.
The mappings and vulnerable methods of lib-vers and CVEs are publicly accessible on our website \cite{dataset}. 

\subsubsection{\textbf{Tools and Environments Preparation}}
All tools used in the evaluation were tested with their latest versions in May 2022. 
\steady was tested with version $3.2.4$ with a built-in vulnerability database including $729$ CVEs. The two commercial tools were evaluated in their publicly accessible production environments. \tool was implemented with 6.9kloc in Python $3.8.2$ and evaluated with
java $7-13$ (depends on projects), Maven $3.8.2$, and Ubuntu $18.04.6$.

\subsection{RQ1: Comparison with Other Remediation Tools}
\subsubsection{\textbf{Evaluation Metrics}}
(1) \textit{Vulnerability fixed}: The primary target of remediation is fixing vulnerabilities which are further classified in terms of reachability as remaining reachable CVEs: $Vul_{r}$, remaining unreachable/unknown CVEs: $Vul_{ur}$/$Vul_{uk}$, and total fix: $Fix$. (2) \textit{Compilation}: $Fail_{comp}$. The projects with updated pom files were compiled by Maven to evaluate the correctness of Maven resolution and the compile-time compatibility. (3) \textit{Unit test}: $Fail_{test}$. The affiliated unit tests were run against the remediated projects to evaluate the runtime compatibility. (4) \textit{Supplementary metrics}: The number of upgraded/downgraded libraries, total version span, number of \ma upgrades (\#Major), and development upgrades (\#Dev) were counted for reference. 
\subsubsection{\textbf{Comparison Results}}
\label{sec:rq1}
The evaluation was conducted based on the remediated projects (versions returned by remediation tools were adjusted in pom files), which along with the Maven logs, are available on our website. To emphasize the improvement gained from the version selection strategy, we added two baseline tools with naive strategies. Both baseline tools share the same partitioning and backtracking mechanisms as \tool. \bla always prefers the latest versions of vulnerable libraries. It is used to demonstrate the result of a common practice which is upgrading vulnerable dependencies to the latest. \blb always prioritizes the versions with the fewest reachable and unknown vulnerabilities, even if it may break the projects. \blb gave an idea of how many non-trivial vulnerabilities could be fixed without being constrained by compatibility. The comparison results with remediation tools and baselines are provided in Table~\ref{tab:comparison}. The analysis of each metric is supplied as follows:

\begin{table*}
\footnotesize
\caption{Comparison of \tool among State-of-the-art Remediation Tools}
\setlength{\tabcolsep}{1.5pt}
\begin{tabular}{lrrrrrrrrrrrr}
    \toprule
\textbf{Tool name}  & \multicolumn{1}{l}{\textbf{Avg \depg Size}} & \multicolumn{1}{l}{\boldsymbol{$Vul_{r}$}} &
\multicolumn{1}{l}{\boldsymbol{$Vul_{ur}$}} &
\multicolumn{1}{l}{\boldsymbol{$Vul_{uk}$}} & \multicolumn{1}{l}{\boldsymbol{$Fixed_{ CVE}$}} & \multicolumn{1}{l}{\boldsymbol{$Fail_{comp}$}} & \multicolumn{1}{l}{\boldsymbol{$Fail_{test}$}} & \multicolumn{1}{l}{\textbf{\#Crashes}} & \multicolumn{1}{l}{\textbf{\#Libs changed}} & \multicolumn{1}{l}{\textbf{Version span}} & \multicolumn{1}{l}{\textbf{\#Dev}} & \multicolumn{1}{l}{\textbf{\#Major}} \\
    \midrule
Original        & 33.99                    & 17                                 & 5,363           & 2,954                      & 0                               & 0                                             & 0                                           & 0                                  & 0                                  & 0                                & 0                                 & 0                                   \\
\textbf{\tool}       & \textbf{36.27}                    & \textbf{0}                                  & \textbf{553}   & \textbf{583}                              & \textbf{7,198}(87\%)                            & \textbf{4}                                             & \textbf{15}                                          & \textbf{0}                                  & \textbf{2,556}                               & \textbf{70,464}                            & \textbf{3}                                & 139                                 \\
\dpb & 34.93                    & 16                                 & 5,357             & 2,682                    & 262 (3\%)                             & 20                                            & 31                                          & 1                                  & 602                                & 17,024                            & 0                                 & 44                                  \\

\steady     & 44.17                   & 11(4)                                 & 1,596(955)          & 1,457(515)                       & 5,253(63\%)                            & 27                                            & 36                                          & 1                                  & 2,292                               & 75,380                            & 4                                 & 257                                 \\
\snyk       & 34.24                    & 7                                  & 4,199       & 2,410                          & 1,469 (18\%)                            & 51                                            & 61                                          & 7                                  & 1,398                               & 24,679                            & 0                                 & 245                                 \\
\scantist   & 35.61                    & 3                                  & 1,040              & 1015                   & 6,277(75\%)                            & 54                                            & 70                                          & 0                                  & 6,498                               & 134,407                           & 0                                 & 170 \\
\bla   & 33.81                   & 3                                  & 4,677            & 2,786                     & 869                             & 39                                            & 45                                          & 0                                  & 2,580                               & 16,863                            & 7                                 & 194                                 \\

\blb 	&43.11	&0	&422	&376	&7,536 & 54& 71 &0 & 5,860	& 90,931	& 5	& 329 \\
\blc  &35.11	&0	&535	&547	&7,252	&4	&12	&0	&2,613	&56,738	&1	&126 \\
    \bottomrule

\end{tabular}
\begin{enumerate}
    \item $Vul_{r}$: number of reachable CVEs. $Vul_{ur}$: number of unreachable CVEs. $Vul_{uk}$: number of \textit{unknown} CVEs. $Fixed_{ CVE}$: number of fixed CVEs. $Fail_{comp}$: number of projects with failed compilation. $Fail_{test}$: number of projects with failed tests. \textit{Crashes}: number of projects that tools crashed and failed to return results. \textit{Dev}: number of development version pairs. \textit{Major}: number of \ma version pairs
\end{enumerate}
\label{tab:comparison}
\end{table*}
   
\begin{itemize}[leftmargin=9pt]
    \item \textbf{Remaining Reachable Vuls}: Due to a limited number of vulnerable methods, only $17$ CVEs could be identified as reachable in original projects. It is noteworthy that \tool eliminated all reachable CVEs. Because \dpb returned far fewer remediation suggestions than other tools, $16/17$ reachable CVEs remained reachable after remediation. As \steady's vulnerability database is limited, we re-evaluated \steady with the $729$ CVEs in their database and enclosed the updated numbers in brackets. Within this scenario, \steady had fewer reachable CVEs than before, like other tools. 
    \item \textbf{Remaining Unreachable and Unknown Vuls}: \tool had much fewer unreachable vulnerabilities (reduce $87.56\%$ of vulnerabilities) than other tools because though the unreachable vulnerabilities were considered harmless, \tool attempted to remove them if the constraints allowed. Unknown vulnerabilities $583$ still remained in the \depg for three reasons: (1) $244, 41.87\%$. The versions with fewer vulnerabilities did not satisfy the constraints. (2) $149, 25.56\%$. All versions were vulnerable. (3) $101, 17.31\%$. The more secure versions with unreachable CVEs were \ma upgrades with overly large version spans. Regarding baselines, \bla proved that upgrading to the latest fixed only an insignificant amount of vulnerabilities. \blb suggested that $338$ ($4.05\%$) more vulnerabilities could be fixed without considering compatibility.
    \item \textbf{Compilation Failures}: \tool achieved $98.67\%$ successful compilation rate due to detecting syntactic breaking and DC issues. The reasons for four failed cases were (1) Call graph generation failure: One of the libraries along the call chain had no call edges generated, which led to unreachable breaking methods. (2) Exception class not captured: The breaking exception class was not captured in the call graph and thus deemed unreachable. (3) Overriding not captured: The breaking methods of a class were extended and overridden in the new version, but the call graph did not reflect such overriding. {For example, a project, \textit{apollo-client}~\cite{apollo}, had a failed compilation due to the incompatibility in its dependency, \textit{snakeyaml}. The overriding of class, \textit{BaseConstructor}, was not captured.} (4) Ghost dependency: The breaking methods were used in an undefined dependency, so they were not captured as reachable methods. Because of the local optimization and unreliable or absent compatibility detection, the rest of the tools were subject to broken upgrades with failed compilation. 
    \item \textbf{Unit Test Failures}: Since it is challenging to detect Semantic Breaking effectively, it is difficult to prevent Unit test failures. Thanks to the prioritization based on SemVer and Maven resolution rules, \tool was able to achieve the fewest failures among these tools. Note that due to private dependencies, unfinished development, special requirements of running environments, etc., $88$ unit tests in original projects already failed without remediation, which was excluded from the number of failures in the table. 
    \item \textbf{Other Statistics}: It is evident that \scantist had many more lib-ver pairs changed because it manipulated the direct dependencies to adjust the associated trees by changing the default versions of subsequent dependencies regardless of vulnerabilities, while other tools mostly focused on the vulnerable vertices. The same reason stood for the version span. Because \tool, \steady, and \scantist substantially changed the versions of transitive dependencies, their total version spans were larger than \dpb's.
\end{itemize}

\begin{figure}[!t]
  \includegraphics[width=1\linewidth]{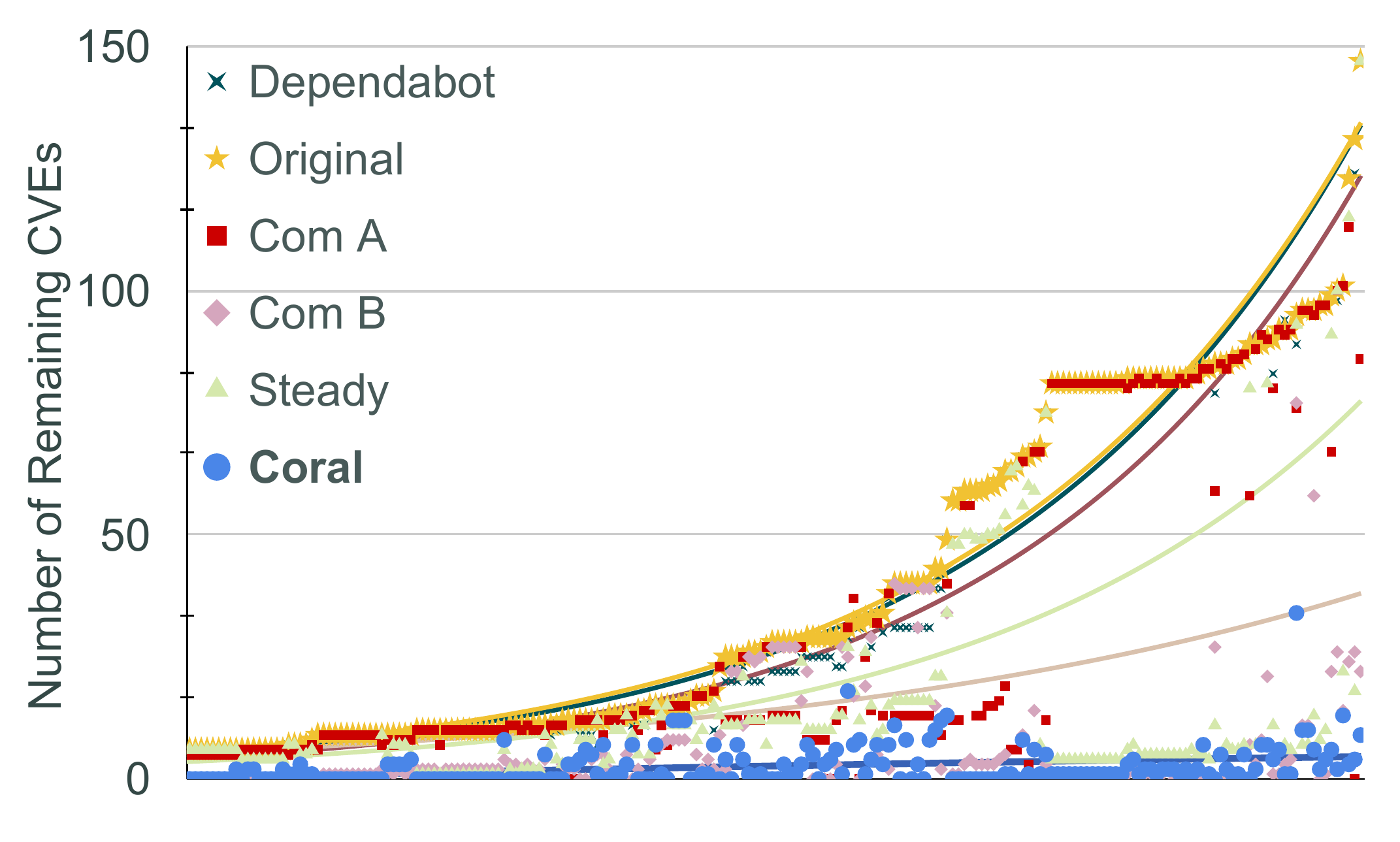}
  \caption{Ascending Order by Numbers of CVEs of Original per Project} 
  \label{fig:vul}
\end{figure}

To illustrate the distribution of remaining vulnerabilities over all projects, the remaining CVEs of all tools are presented in the scatter plot of Fig.~\ref{fig:vul}. The x-axis is ordered by the number of CVEs of the original, which serves as the upper bound denoted by yellow stars. It is evident that \tool has the overall fewest remaining CVEs at the bottom of the chart, denoted by blue dots.

\begin{tcolorbox}[size=title,opacityfill=0.2,breakable]
\textbf{Conclusions of RQ1: }From the evaluation in Table~\ref{tab:comparison}, \tool fixed $87.56\%$ of all CVEs with all reachable removed, including $911$ more CVEs than the best of the rest tools. Meanwhile, \tool achieved the $98.67\%$ successful compilation rate and $92.96\%$ successful unit test rate, which outperformed the rest of the tools. Compared with the two baseline tools, \tool was proven to be effective at balancing the compatibility and security by breaking $106$ ($35.21\%$) fewer projects at the cost of $338$ ($4.05\%$) fewer vulnerabilities fixed. 
\end{tcolorbox}

\subsection{RQ2:Effectiveness of Improvement on Global Optimization}
\begin{figure}[!t]
  \includegraphics[width=1\linewidth]{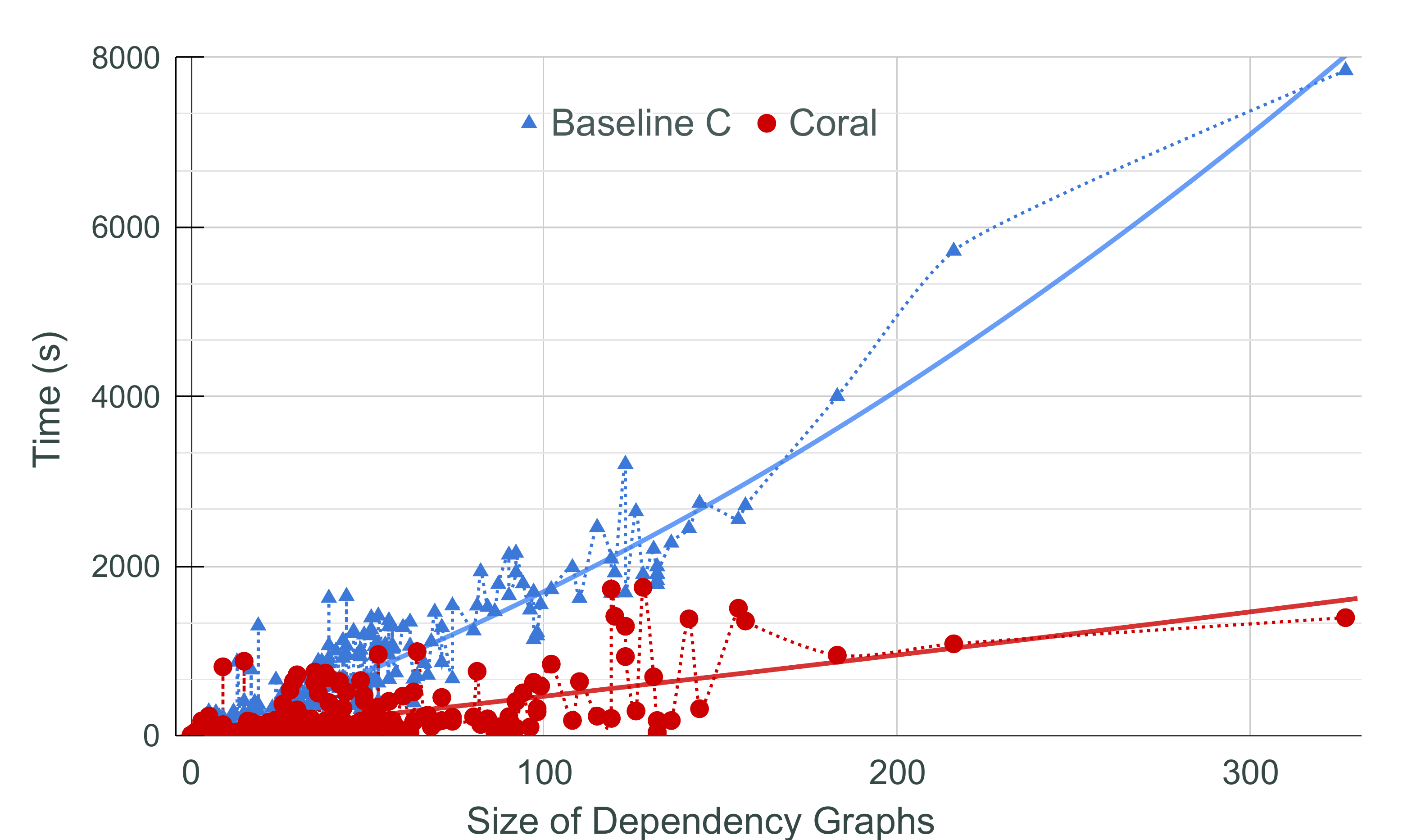}
  \caption{Time Consumption of \blc and \tool} 
  \label{fig:rq2}
\end{figure}

The subgraph partitioning was implemented in \tool to boost the performance towards the global optimization. To evaluate the effectiveness of partitioning, \blc was implemented in the same logic without two types of partitioning used by \tool. The same data set was used to evaluate the existing metrics and time consumption. To measure the time, we respectively ran \tool and \blc ten times against each project and calculated the average time as the final result. 
The result is presented in Fig.~\ref{fig:rq2}, which illustrates that the \blc generally tended to spend more time than \tool for complete remediation. In the figure, the $301$ projects are ordered by the size of \depg. Each dot in the figure represents a single project. The tendency curves of both are fitted by the second-degree polynomials to avoid over-fitting. 

To explain the fluctuations of the consumed time of \tool, we manually analyzed the causes of the outliers. First, the $6$ lower outliers were collected and analyzed. The cause of these cases was subgraphs partitioned were pretty small (1-5 deps), and the backtracking was not triggered. Second, for $18$ higher outliers, there were four major causes:
\begin{itemize}[leftmargin=9pt]
\item (9 cases) \textbf{Call graph generation failures}: The Call graph generation of the Soot script failed at some dependencies of \depg, which took a long time to return. Usually, the failure of one version would persist with other versions of the same library, so the total time was elongated. 
\item (5 cases) \textbf{SMT solver took a long time}: For these projects, both \blc and \tool spent a long time because the SMT Solver took a long time to finish. The direct reason for this cause was that the levels of \depgs were limited, which means the \depgs were more flattened than others. Thus, the partitioning of \tool based on levels could still include a substantial number of vertices in the SMT solver. 
\item (3 cases) \textbf{Multiple backtracking}: Another backtracking could be triggered during the current run of backtracking or after the current run fails. In these cases, 2 out of 3 cases had over three attempts of failed hard backtracking, and the rest triggered the hard backtracking multiple times during soft backtracking, which led to no improvement of vulnerability reduction for this soft backtracking. 
\item (1 case) \textbf{Jar downloading failure}: The CG generation and Syntactic breaking detection relied on the jar files of dependencies, \tool failed to download from Maven Repository with time-out multiple times. 
\end{itemize}

The observed metrics for \blc are presented in Table~\ref{tab:comparison}. From the table, \blc has fixed $54$ ($0.75\%$) more vulnerabilities than \tool, which implies the global optimization without partitioning has slightly improved the vulnerability fixing. Moreover, the number of projects with failed compilation stayed the same because \tool handled the syntactic breaking and DC issues regardless of the partitioned subgraphs by backtracking.

\begin{tcolorbox}[size=title,opacityfill=0.2,breakable]
\textbf{Conclusions of RQ2: }
The comparison between \blc and \tool substantiates that the partitioning mechanism could substantially reduce the time consumption without introducing the compilation failures at an acceptable cost of $0.75\%$ fewer fixed vulnerabilities, especially for the large \depg.
\end{tcolorbox}

\subsection{RQ3: How many fixable/unfixable CVEs in Maven}
We target finding out how many vulnerabilities can be fixed without breaking the compilation and how many cannot in popular Maven projects. Since \tool could efficiently exclude the solutions that broke the compilation with high precision ($98.67\%$), we made an assumption that CVEs fixed by \tool were fixable and CVEs not fixed by \tool were unfixable.
\subsubsection{\textbf{Preparation of
data}} To conduct a large-scale study in the Maven ecosystem, we constructed a different data set from RQ1 and RQ2. {Considering the balance between the representativeness and quality of the dataset,} we first collected repositories with 100+ stars managed by Maven from GitHub to ensure the high quality of the dataset. Then, we compiled them and extracted dependency trees from them by the Maven command. If both steps succeeded, the dependency trees and class files were used as input for the remediation. Eventually, we randomly selected $2,000$ out of $6,898$ projects (average size $103.58$) for the evaluation {to make sure the dataset was representative}. As for CVE mappings, the same mappings were used as RQ1 and RQ2. Since collecting vulnerable methods and classes is not as straightforward as version mappings, which requires much more effort for all CVEs, we decided not to conduct the reachability analysis of vulnerabilities in the experiment.

\subsubsection{Results of RQ3}
\begin{figure}[!t]
  \includegraphics[width=1\linewidth]{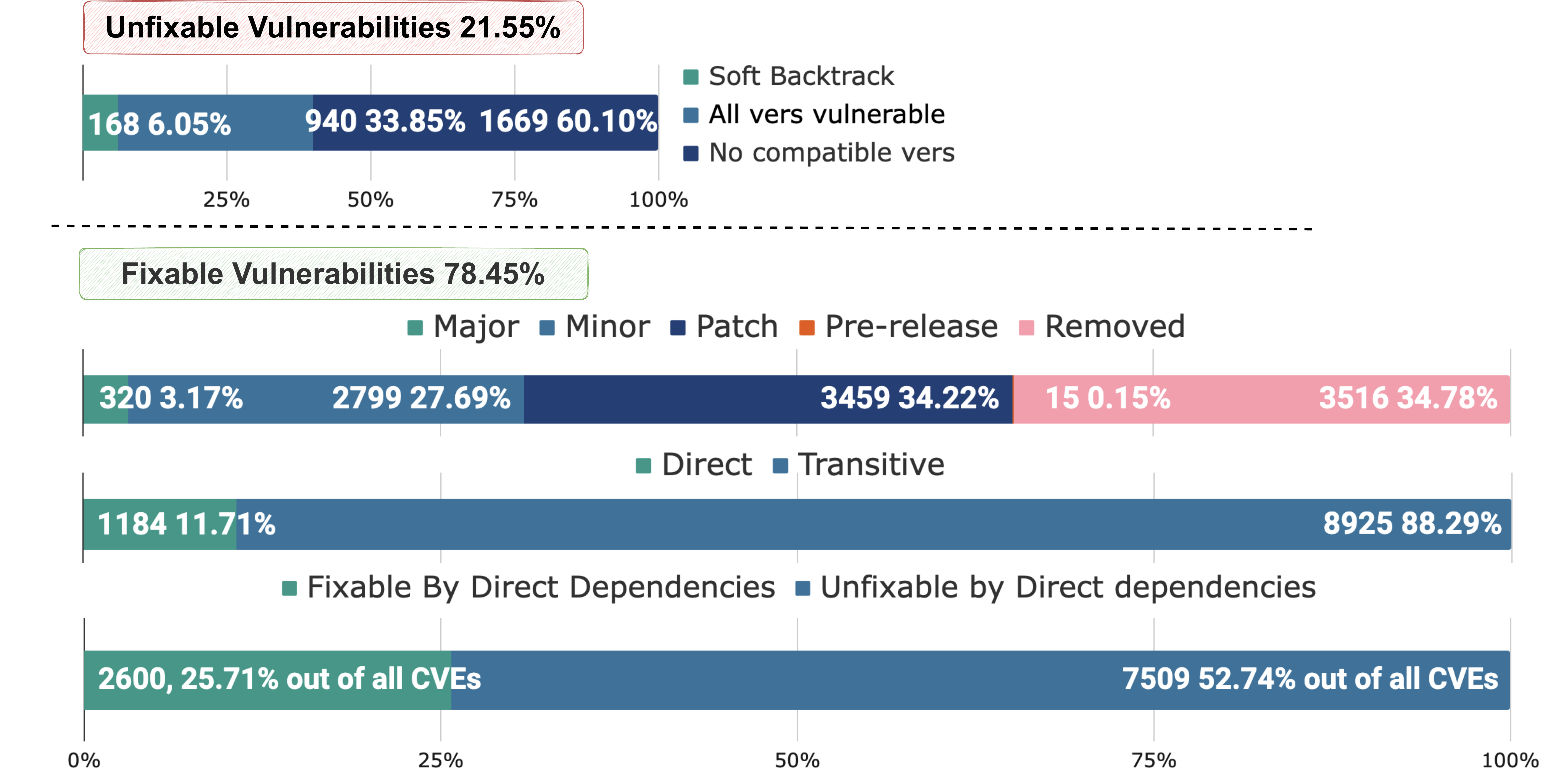}
  \caption{Distributions of Fixable/Unfixable Vulnerabilities} 
  \label{fig:rq3}
\end{figure}

\textbf{Fixable:} The fixable CVEs are $10,109$ ($78.45\%$) as in Fig.~\ref{fig:rq3}. It is inferred that around $78\%$ vulnerabilities could have been safely eliminated from the popular Maven projects without breaking the compilation to reduce the vulnerability risks of the ecosystem. We further calculated the distribution of the CVEs regarding the levels of the libraries and the types of upgrades that removed the CVEs. Although $78\%$ seems to be a large number, the majority of them could not be fixed without domain knowledge or the aid of \tool. According to Fig.~\ref{fig:rq3}, the proportion of vulnerabilities that could be fixed by adjusting direct dependencies was $11.71\%$, out of which $8.34\%$ belonged to \mi and \pa upgrades. 

As users can straightforwardly upgrade their direct dependencies to non-major versions to fix vulnerabilities on their own, we applied this naive method for the comparison with \tool. The result showed that $25.71\%$ of CVEs can be fixed by upgrading direct dependencies. Due to \rip, not only were $8.34\%$ in direct dependencies fixed but more CVEs in transitive dependencies were also fixed. It is implied that without the aid of \tool, the rest of the fixable vulnerabilities ($52.74\%$) could not be fixed straightforwardly. 

\textbf{Unfixable}: The number of unfixable CVEs was $2,777$ ($21.55\%$) as in Fig.~\ref{fig:rq3}, which could not be fixed by \tool for three major reasons, the soft backtrack, all versions of a library were vulnerable, and the secure versions were incompatible. Reflected in Fig.~\ref{fig:rq3}, it is observed that the major reason was incompatibility which accounted for $60.10\%$. Note that the incompatibility did not count the Semantic Breaking because it could not be reliably detected. The minor reason, soft backtrack, refers to the vulnerabilities being left unfixed because the soft backtrack could not eradicate all CVEs, but minimized the overall vulnerabilities by ignoring some CVEs.

Although unfixable vulnerabilities cannot be easily removed without breaking the projects, some of them are removable at an acceptable cost. For example, if an API is deprecated and migrated to another, users only have to invoke the updated API and upgrade to the target version to fix the issue and vulnerabilities. Thus, if efforts to fix incompatibility are acceptable, more vulnerabilities can be fixed thoroughly with minimized efforts by quantifying efforts to fix the incompatibility.

\begin{tcolorbox}[size=title,opacityfill=0.2,breakable]
\textbf{Conclusions of RQ3:}
Through the experiments with the most starred projects on GitHub, we found that $78.45\%$ of vulnerabilities could be fixed without breaking the compilation. However, without the aid of \tool, only $25.71\%$ could be fixed by upgrading the direct dependencies to non-major secure versions. 
\end{tcolorbox}

\section{Threats of Validity}
The threat of \tool is the static call graph reliance because only the static call graphs are not accurate enough to capture all possible call edges, which is one of the causes of the unit test failures in \Cref{sec:rq1}. {One typical example of inaccurate static call graphs is that static call graphs may miss invocations made by dynamic features, e.g., reflection.} Moreover, The prioritization of vulnerabilities might overlook some reachable ones due to inaccurate static call graphs. However, to modularly and dynamically update the call graphs after each version adjustment, we could only generate static call graphs that are faster than dynamic ones. As it is impractical to run tests and generate dynamic call graphs thousands of times per project, we sacrificed accuracy for better performance.
\section{Related Work}

\subsection{SCA Tools and Strategies}
SCA has been a popular research topic in recent years. Researchers have invested much effort to study and improve the two major procedures: component and vulnerability detection and vulnerability remediation. 

\subsubsection{\textbf{SCA remediation}}
A limited number of research works \cite{alfadel2021use,pashchenko2020vuln4real,ponta2018beyond,ponta2020detection,soto2021longitudinal} attempted to study and enhance the remediation strategy.
Alfadel et al. \cite{alfadel2021use} found for the Javascript projects at Github $34.58\%$ of PRs created by \dpb were not merged 
due to five reasons: (1) Duplication (2) Dependency conflict by peer requirements (3) Test failures (4) Internal errors (5) Disobeying rules/standards, which substantiate our findings in Section \ref{sec:mavenstudy}.
\steady \cite{ponta2018beyond, ponta2020detection} has been developed for years to be a code-centric and usage-based SCA tool, which has been proved effective by Imtiaz et al. \cite{imtiaz2021comparative}.
Soto et al. \cite{soto2021longitudinal} found $22.6\%$ of upgrades by \dpb were recommended
for bloated dependencies. $22.6\%$ does not
contradict our result $3.92\%$ because $22.6\%$ consists of all
bloated dependencies, while ours were only those found and addressed.
These works except for \steady mostly focused on the evaluation of remediation tools, which left a blank of remediation strategy enhancement filled by \tool.

\subsubsection{\textbf{Component and vulnerability detection}}
Many researchers and practitioners \cite{imtiaz2021comparative,dann2021identifying, alfadel2021empirical,pashchenko2020vuln4real,pashchenko2018vulnerable,owasp,blackduck,whitesource,sourceclear,sonarqube,zhang_compatibility, zhan2021atvhunter,zhan2020automated} have studied the component and vulnerability detection. Imtiaz et al. \cite{imtiaz2021comparative} studied 9 commercial SCA tools and found the reported vulnerabilities vary substantially, which revealed that the vulnerability database  was the key differentiator. Dann et al. \cite{dann2021identifying} reviewed six commercial and academic SCA tools regarding their ability to handle the dependency modification types. By testing 7k+ Java projects, they found the re-bundle modification in Maven dependencies was not supported by any tools. 
\textit{Vuln4real} \cite{pashchenko2020vuln4real,pashchenko2018vulnerable} was proposed to exclude the false alarms of vulnerabilities by identifying the vulnerabilities in lagging, development-only, and unreachable dependencies, which significantly reduces false alerts. 

\subsection{Study of Open-source Software Ecosystem}
Apart from SCA techniques, researchers \cite{decan2018impact, imtiaz2022open, ponta2019manually, perl2015vccfinder, alqahtani2016tracing,liu2022demystifying,hejderup2015dependencies,li2016vulpecker,farris2018vulcon, plate2015impact, tang2022towards} have studied the open-source software (OSS) and associated vulnerabilities in the OSS ecosystem, conclusions of which can be used to guide the designs of SCA tools. Decan et al. \cite{decan2018impact} studied NPM and Rubygems package managers and found that $33\%$
and $40\%$ of vulnerabilities respectively had their fixes within the same
major release.
Plate et al. \cite{plate2015impact} proposed new metrics to determine the criticality of vulnerabilities regardless of the types and languages of vulnerabilities, which helps with the automated impact assessment of  new vulnerabilities.  
Imtiaz et al. \cite{imtiaz2022open} studied the characteristics of security fixes at 6 major package managers, namely, the semantic versions, release notes, and the time lag between fixes and releases, and offered 4 recommendations for the better practice of security releases. Ponta et al. \cite{ponta2019manually} manually collected $625$ publicly disclosed vulnerabilities for Java projects, which was also used in the \Cref{sec:evaluation} as the \steady data set at the latest version.
\section{Conclusion}
We proposed \tool to provide remediation without breaking compatibility. The evaluation demonstrated that \tool outperformed other tools by fixing $87.56\%$ of vulnerabilities and achieving $98.67\%$ successful compilation rate and $92.96\%$ successful unit test rate. In the ablation study, the partitioning of \depg and trade-off between security and compatibility had been proved effective. Furthermore, we found that $78.45\%$ of vulnerabilities in popular Maven projects could be fixed without breaking the compilation.

\section*{Data Availability}
The data sets of the studies and evaluations can be publicly accessed at 
\textit{https://sites.google.com/view/icse23remediation}.

\section*{Acknowledgements}
\label{sec:ack}
This research is partially supported by the National Research Foundation Singapore and DSO National Laboratories under the AI Singapore Programme (AISG Award No: AISG2-RP-2020-019), the NRF Investigatorship NRF-NRFI06-2020-0001, the National Research Foundation through its National Satellite of Excellence in Trustworthy Software Systems (NSOE-TSS) project under the National Cybersecurity R\&D (NCR) Grant award no. NRF2018NCR-NSOE003-0001, the Ministry of Education, Singapore under its Academic Research Fund Tier 2 (MOE-T2EP20120-0004) and Tier 3 (MOET32020-0004). Any opinions, findings and conclusions or recommendations expressed in this material are those of the author(s) and do not reflect the views of the Ministry of Education, Singapore.

\clearpage
\balance
\bibliographystyle{IEEEtran}
\bibliography{acmart}

\end{document}